\documentclass[aps,prl,floatfix,twocolumn,superscriptaddress]{revtex4}
\usepackage{amsmath,amssymb}
\usepackage{graphicx}
\usepackage{epsfig}
\usepackage[normalem]{ulem} % ??? ??????????? ??????
\usepackage[usenames]{color}

\usepackage[T1]{fontenc}
\usepackage[x11names]{xcolor}
\usepackage{empheq}

\begin{document}

%\twocolumn
%\hsize\textwidth\columnwidth\hsize\csname@twocolumnfalse\endcsname

%\title {Resonant Photon Drag of dipolar excitons in normal and Bose-condensed regimes}

%\title {Resonant Circular Photon Drag Effect in a Bose-Einstein Condensate}

%\title{Frequency-Selective Photodetector Based on the Resonant Photon Drag Effect\\ in a Bose--Einstein Condensate}

\title{Proposal for frequency-selective photodetector based on the resonant photon drag effect in a condensate of indirect excitons}

\author{V.~M.~Kovalev}
%\email{vadimkovalev@isp.nsc.ru}
\affiliation{A.V. Rzhanov Institute of Semiconductor Physics, Siberian Branch of Russian Academy of Sciences, Novosibirsk 630090, Russia}
\affiliation{Department of Applied and Theoretical Physics, Novosibirsk State Technical University, Novosibirsk 630073, Russia}

\author{M.~V.~Boev}
%\email{vadimkovalev@isp.nsc.ru}
\affiliation{A.V. Rzhanov Institute of Semiconductor Physics, Siberian Branch of Russian Academy of Sciences, Novosibirsk 630090, Russia}

\author{I.~G.~Savenko}
\affiliation{Center for Theoretical Physics of Complex Systems, Institute for Basic Science (IBS), Daejeon 34126, Korea}

\date{\today}
\begin{abstract}
We present a microscopic theory of a photon drag effect that appears in a Bose--Einstein condensate of neutral particles, considering indirect excitons in a double quantum well nanostructure under the action of a polarized electromagnetic field. It is shown that the dynamical polarization of excitons results in a resonant behavior of the exciton photon drag flux when the frequency of light is close to the gap between two energy levels of internal exciton motion. Specifically, we consider the ground and first excited energy states characterized by the angular momentum difference $\pm 1$, and thus, the helicity of light matters. We show that the resulting drag current is caused by both Bose-condensed particles and the particles in the excited states. As a result, the total current represents a superposition of thresholdlike and resonant contributions,--- property, which can be used in frequency-selective photodetection.
\end{abstract}

\maketitle

%\newpage

%------------------
%------------------
%------------------

\textit{Introduction. }The photon-drag effect (PDE) owes its appearance to photon translational momentum, and it serves as a paramount manifestation of radiation pressure~\cite{Gibson, Danishevskii, Ivchenko}. Conventionally, when atoms in a gas~\cite{Werij} or free carriers of charge, such as conduction band electrons and valence band holes, absorb radiation while interacting with electromagnetic waves, they start to move in a direction predefined by the momentum of the photons. This effect has been widely studied in semiconductors~\cite{Costa}, dielectrics~\cite{Loudon}, metals~\cite{Goff}
including thin films~\cite{Kurosawa} and bulk tellurium~\cite{Shalygin}, monolayer and multilayer graphene~\cite{Glazov, Entin}, carbon nanotubes~\cite{Mikheev}, topological insulators~\cite{Plank}, semiconductor microcavities~\cite{Berman}, and two-dimensional electron gas~\cite{Wieck, Luryi, Grinberg}, where the first theories of the PDE were developed for Germanium and polar crystals in a model based on electron--photon--acoustic phonon interaction~\cite{Grinberg2, Yee}.

One of the possible extensions of the PDE is the circular photon drag effect (CPDE), which occurs when light simultaneously transfers both translational and angular momenta to electrons~\cite{Belinicher} (or holes). This effect manifests itself in the helicity-dependent photon drag current of light~\cite{Keller, Shalygin2}, observed in multiple two-dimensional systems such as metamaterials and graphene~\cite{Karch, Kang}.

It is important to specify that interaction of electrons and holes with an electromagnetic field (EMF) is due to the presence of electric charge. 
The situation is more complicated in the case of neutral particles, such as atoms or excitons. 
Due to absence of charge, their interaction with the EMF is usually much weaker. However, in some cases the light pressure can reach significant values. This phenomenon is referred to as the \textit{resonant light pressure}~\cite{Kazansev}.
Qualitatively, it can be explained as follows. When a neutral particle is exposed to an EMF, it acquires an induced dipole moment, which in the linear limit (of not very strong fields) has the structure $p=\alpha(\omega)E$, where $\alpha(\omega)$ is the dynamical polarizability of the atom. The energy of interaction of the dipole moment with the EMF reads $U=-pE$. As a result, the particle experiences the influence of constant force $F=-\nabla U=\alpha(\omega)\nabla\langle E^2\rangle$, where the angular brackets stand for the time average. Clearly, this force strongly depends on frequency through $\alpha(\omega)$. Indeed, as follows from a standard perturbation theory analysis, the tensor of polarizability reads~\cite{Landau8}
\begin{eqnarray}
\nonumber
\alpha_{ij}(\omega)=\sum_{n}\left(\frac{(d_i)_{gn}(d_j)_{ng}}{\omega_{ng}-\omega}
+
\frac{(d_j)_{gn}(d_i)_{ng}}{\omega_{ng}+\omega}\right),
\end{eqnarray}
where $\hat{d}$ is the operator of dipole moment of the atom, and $\omega_{ng}=E_n-E_g$ is the energy difference between the ground $|g\rangle$ and excited $|n\rangle$ states. Due to the resonant character of the polarizability, the force acting on the particle can dramatically increase at various frequencies $\omega\sim\omega_{ng}$ when such transitions are allowed by the selection rules (when the matrix element of the transistion dipole moment is nonzero). This mechanism explains the physics of resonant light--matter interaction.

An analogous effect can take place if we apply the upcoming arguments to quasiparticles in solid state systems. For instance, indirect excitons, which represent Coulomb coupled electron--hole pairs where electrons and holes reside in separate parallel layers, have a discrete energy spectrum and possess a dipole moment. They can be resonantly coupled to light at normal incidence to the structure; moreover, Bose--Einstein condensation (BEC) and superfluidity have been experimentally observed in these systems~\cite{Butov, Fogler}. Beside fundamental interest, indirect excitons can be used in various applications based on the Hall effect~\cite{Arnardottir}, room-temperature transport~\cite{Fedichkin}, hybrid Bose--Fermi systems~\cite{OurFano}, and ballistic spin transport~\cite{Kavokin1}. It should be noted, however, that since excitons are neutral particles, the resonant light pressure results in a particle current rather then electric current in the system. In what follows we will call the particle current caused by the resonant light pressure the \textit{resonant photon drag effect} (RPDE). 

Recently by some of the authors of this manuscript it was suggested to use the phenomenon of radiation pressure to quantize the system response in the presence of a BEC~\cite{RefRPQ}. 
In this article, we describe the RPDE in a BEC of neutral particles represented by dipolar indirect excitons. We show that RPDE of excitons results in a particle current that possesses a number of intriguing peculiarities.

%------------------------------------------------------------------------
%------------------------------------------------------------------------
%------------------------------------------------------------------------

\textit{System and Hamiltonian. }We start from the Hamiltonian describing a single exciton in a double quantum well (DQW) interacting with an EMF:
\begin{gather}\label{eq.1}
H=\frac{\left(\textbf{p}_h-\frac{e}{c}\textbf{A}_h\right)^2}{2m_h}+\frac{\left(\textbf{p}_e+\frac{e}{c}\textbf{A}_e\right)^2}{2m_e}+
U_c(\textbf{r}_e-\textbf{r}_h).
\end{gather}
Here, $m_{e(h)}$ is the effective mass of the electron (hole) and $e>0$ is the hole charge. $U_c(\textbf{r}_e-\textbf{r}_h)$ is electron--hole interaction energy, and $\textbf{A}(\textbf{r},t)=\textbf{A}_0e^{\textrm{i}\textbf{kr}-\textrm{i}\omega t}+\textbf{A}_0^*e^{-\textrm{i}\textbf{kr}+\textrm{i}\omega t}$ is the EMF vector potential, where $\textbf{k}$ is the in-plane component of the EMF wave vector $\textbf{Q}$ (see Fig.~1).
\begin{figure}[!t]
\includegraphics[width=1.0\linewidth]{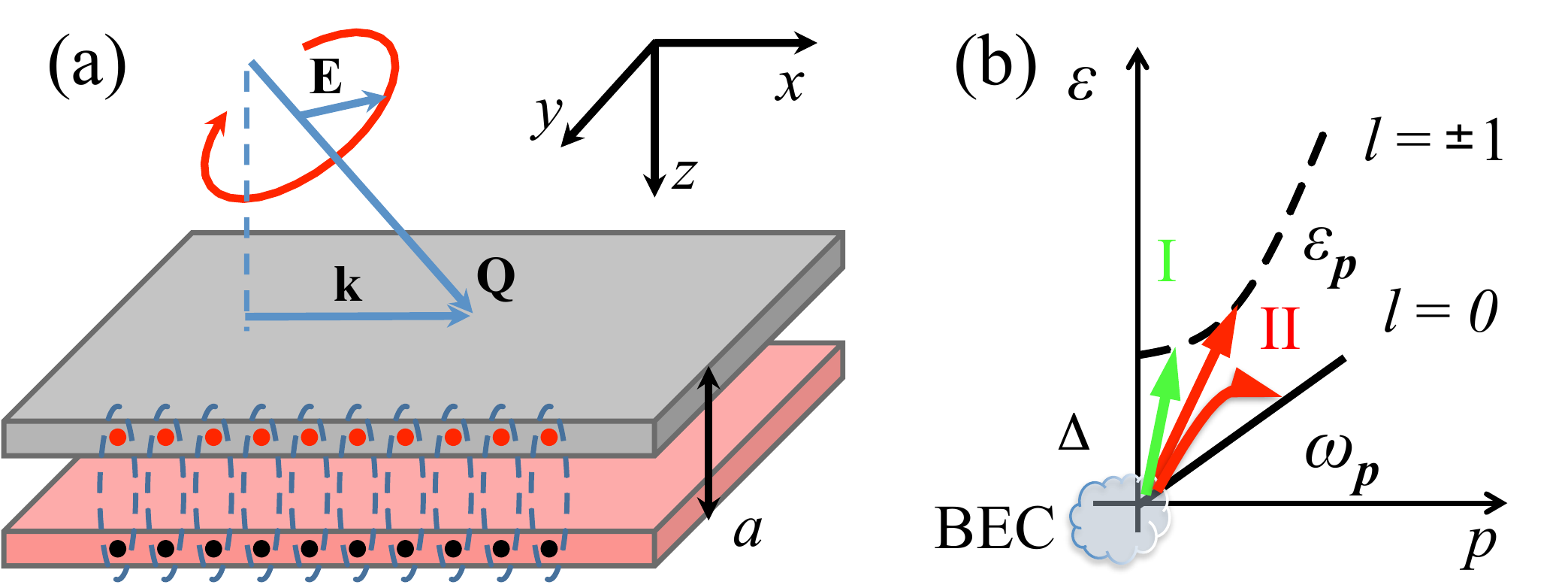}
	\caption{(a) Dipolar exciton gas in an external electromagnetic field. (b) Energy spectrum of excitons in the BEC regime. $l$ indicates the angular momentum of the internal exciton motion, $\varepsilon_{\textbf{p}}$ is the kinetic energy of the exciton center of mass, and $\omega_{\textbf{p}}$ is a Bogoliubov quasiparticle dispersion. Numbers I (green) and II (red) characterize the two principal processes of light absorption.}
\label{Fig1}
\end{figure}
Introducing the relative $\textbf{r}=\textbf{r}_e-\textbf{r}_h$ and center-of-mass $\textbf{R}=(m_e\textbf{r}_e+m_h\textbf{r}_h)/M$ coordinates, and the corresponding momentum operators $\textbf{p}=-\textrm{i}\partial_\textbf{R}$ and $\textbf{q}=-\textrm{i}\partial_\textbf{r}$ (thus $\textbf{p}_e=m_e\textbf{p}/M-\textbf{q}$ and $\textbf{p}_h=m_h\textbf{p}/M+\textbf{q}$) where $M=m_e+m_h$ is full exciton mass, we can rewrite the Hamiltonian~\eqref{eq.1} in the dipole approximation as $H=H_0+U_c(\textbf{r})+U$, where
\begin{gather}
\label{eq.4}
H_0=\frac{\textbf{p}^2}{2M}+\frac{\textbf{q}^2}{2m};~~
U=-\frac{e}{m c}\textbf{q}\textbf{A}+\frac{e^2}{2m c^2}\textbf{A}^2.
\end{gather}
Here, $m^{-1}=m_e^{-1}+m_h^{-1}$ is the reduced exciton mass, and from now on $\textbf{A}=\textbf{A}(\textbf{R},t)$ acts on the center-of-mass dynamics of the exciton only.
In Eq.~\eqref{eq.4}, $H_0+U_c(\textbf{r})$ describes both the center-of-mass exciton motion and the relative motion of the electron and hole constituting the exciton. 

The general form of electron--hole interaction energy potential, $U_c(\textbf{r})$, depends on the type of semiconductor alloy constituting the DQW. While the most widespread DQW employed to study dipolar exciton BEC is based on a GaAs/AlGaAs heterostructure, it should be noted that recently, double-layered van der Waals heterostructures based on dichalcogenide monolayers have become popular, especially in view of dipolar exciton BEC. In particular, the critical condensation temperature in such structures is predicted to reach 300 K.

The potential $U_c(\textbf{r})$ has a general property which plays a crucial role independent of its particular form: it is axially symmetric. Thus, the wave functions of internal exciton motion are characterized by the angular momentum quantum number $l$ and the radial quantum number $n$, with the eigenenergies of internal motion $\varepsilon_{n,l}$ and eigenstates $|n, l\rangle$. Total exciton energy is then $E_{n,l}(\textbf{p})=\varepsilon_{n,l}+\varepsilon(\textbf{p})$, where $\varepsilon(\textbf{p})=\textbf{p}^2/2M$ is exciton center-of-mass kinetic energy. 
The nonzero matrix elements of momentum $\textbf{q}$ couple the states in which the angular momentum quantum numbers differ by $\pm 1$. In particular, if an indirect exciton is initially in the ground state, $|0,0\rangle$, transitions to states $|n,\pm 1\rangle$ are possible. 
Hereafter, we will consider the optical transitions between the $|0,0\rangle$ and $|0,\pm 1\rangle$ states with corresponding energy difference $\Delta=\varepsilon_{0,\pm1}-\varepsilon_{0,0}$, assuming that field frequency $\omega$ is close to this energy ($\omega\approx\Delta$). 
For further referring, we introduce the notations (see Fig.~\ref{Fig1}b):
\begin{gather}\label{eq.5}
\varepsilon_1(\textbf{p})=\varepsilon_{\textbf{p}},\,\,\,
\varepsilon_2(\textbf{p})=\Delta+\varepsilon_{\textbf{p}}.
\end{gather}
\begin{figure}[!b]
	\includegraphics[width=1.0\linewidth]{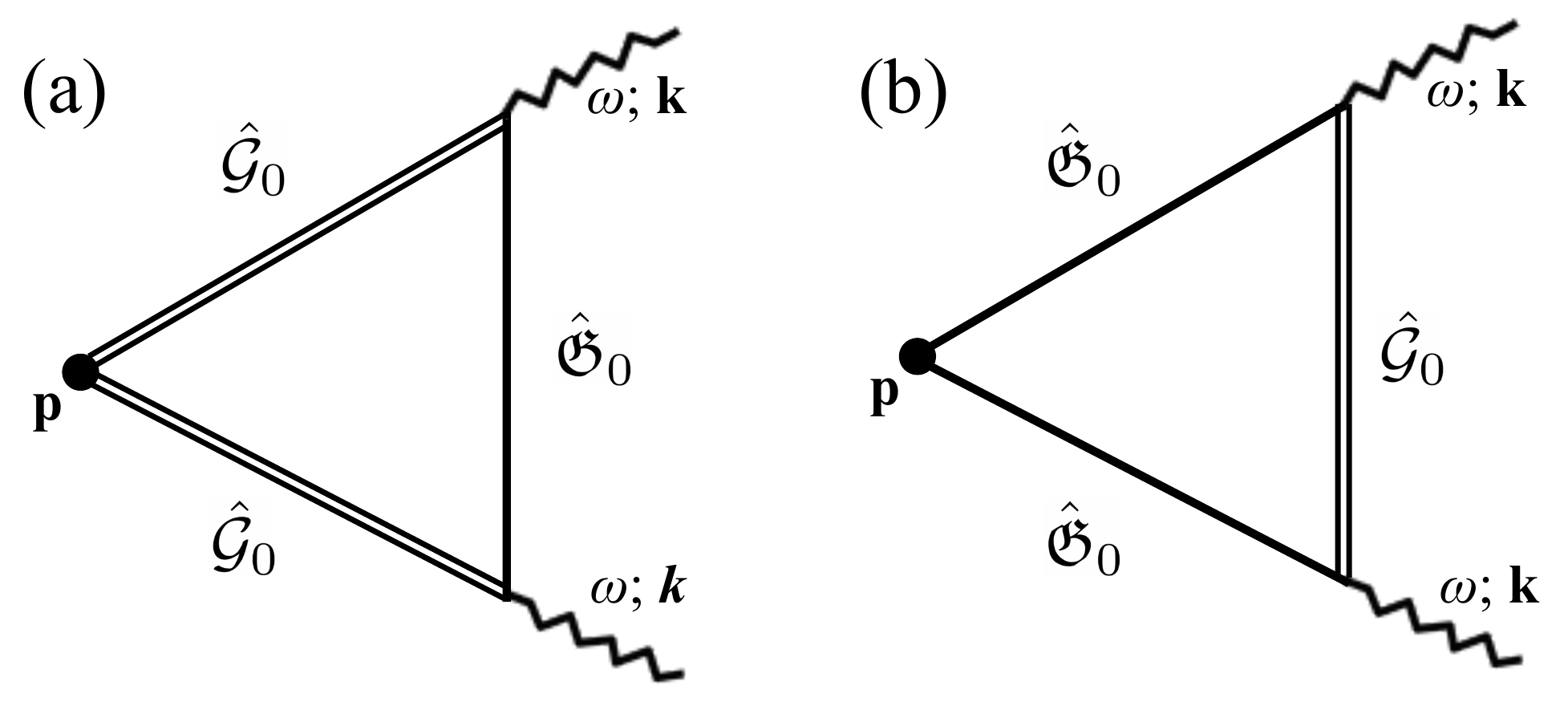}
	\caption{Feynman diagrams describing the components of current (a) $j_c$ and (b) $j_2$. The Green's functions (straight lines) are given in Eq.~\eqref{eq.21w}, wiggling lines stand for the photons, see text for details.}
\label{Fig2}
\end{figure}
%
%
%

%-------------------------------
%-------------------------------
%-------------------------------
%
%
%

%
%
%
\begin{figure*}[!t]
\includegraphics[width=0.99\linewidth]{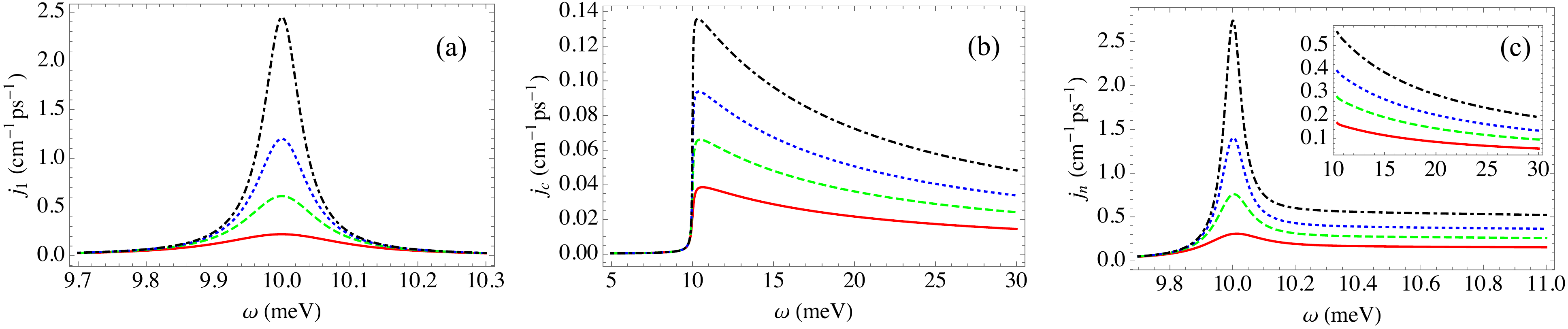}
	\caption{Components $j_c$ (a), $j_1$ (b), and $j_n=j_1+4j_c$ (c) of the current as functions of external EMF frequency for various impurity scattering times: $\tau=2$ ps (red), $5$ ps (green), $7$ ps (blue), and $10$ ps (black). Inset in (c) demonstrates current decay at high $\omega$.}
\label{Fig3}
\end{figure*}

\emph{Photon drag current. }Now we consider a zero-temperature limit, where exciton BEC forms in the lowest exciton state. 
We assume an initial equilibrium condensate at $\varepsilon_1(\textbf{p}=0)$ and an empty excited state at $\varepsilon_2(\textbf{p})$.
We also assume that the external electromagnetic wave excites the excitons from the condensate to $\varepsilon_2(\textbf{p})$, with an exciton density at this level much less than BEC density $n_c$. 

Consequently, we disregard exciton--exciton interaction between the BEC and excited excitons, thereby keeping only the exciton--exciton interaction in the BEC. Thus, the exciton system can be described by the two-component spinor wave function ($x=(\textbf{R},t)$):
\begin{gather}\label{eq.15}
\Psi(x)=\left(
         \psi^*_1(x),\psi^*_2(x) 
     \right)^T,
\end{gather}
which satisfies the equation
\begin{gather}\label{eq.16}
\left(
  \begin{array}{cc}
    \textrm{i}\partial_t-\varepsilon_1(\textbf{p})+\mu-g|\psi_1|^2 & \frac{e}{m c}\textbf{q}_{12}\cdot\textbf{A} \\
    \frac{e}{m c}\textbf{q}_{21}\cdot\textbf{A} & \textrm{i}\partial_t-\varepsilon_2(\textbf{p}) \\
  \end{array}
\right)
\Psi(x)=0,
\end{gather}
where $\mu$ is exciton BEC chemical potential, and $g$ is exciton--exciton interaction strength.
Here we disregard the terms $\sim\textbf{A}	^2$ which also appear in Eq.~\eqref{eq.4}. 

We take the external electromagnetic field to be circularly polarized, and thus it can be characterized by the polarization degree $\sigma=\pm1$. Then the matrix elements of the momentum of relative motion $\textbf{q}_{12}$ can be calculated over the ground and excited exciton states as $\textbf{q}_{12}=\langle0,0|\textbf{q}|0,\pm1\rangle$. They can be expressed through the matrix elements of the internal exciton coordinate operator as $\textbf{p}_{12}=i\Delta m\textbf{r}_{12}$.  
%%%

Exciton RPDE represents exciton flux, which appears in the second order of the external field $\textbf{A}(x)$, and can be characterized by surface density with a dimension of ps$^{-1}$cm$^{-1}$. It can be found by time-averaging the standard quantum-mechanical expression
\begin{gather}\label{EqCurrent}
\textbf{j}=\frac{\textrm{i}}{2M}\sum_{i=1,2}\langle\psi_i\nabla_{\textbf{R}}\psi_i^*-\psi_i^*\nabla_{\textbf{R}}\psi_i\rangle_{t},
\end{gather}
where $i=1$ corresponds to the contribution of the BEC [$\psi_1(x)$ component of the spinor~\eqref{eq.15}], and $i=2$ is the contribution of the excited states [$\psi_2(x)$].

Considering here the EMF, $\textbf{A}(x)$, as a perturbation, we can replace $\psi_1(x)\rightarrow\psi_0+\delta\psi_1(x)$ and $\psi_2(x)\rightarrow\delta\psi_2(x)$, where
$\psi_0$ describes the BEC state, with $n_c=|\psi_0|^2$. Linearizing Eq.~\eqref{eq.16}, we find the following system of equations:
\begin{eqnarray}\label{eq.20w}
\hat{\mathcal{G}}^{-1}_0\delta\hat{\psi}_1(x)&=&-\frac{e}{mc}\textbf{A}(x)\hat{\textbf{q}}\delta\hat{\psi}_2(x),
\\
\label{EqPsi2}
\hat{\mathfrak{G}}^{-1}_0\delta\hat{\psi}_2(x)&=&-\frac{e}{mc}\textbf{A}(x)\hat{\textbf{q}}^*\Bigl(\hat{\psi}_0+\delta\hat{\psi}_1(x)\Bigr),
\end{eqnarray}
where
\begin{gather}\label{eq.21w}
\hat{\textbf{q}}=\left(
                     \begin{array}{cc}
                       \textbf{q}_{12} & 0 \\
                       0 & \textbf{q}_{12}^* \\
                     \end{array}
                   \right),\,\,
\delta\hat{\psi}_i(x)=\left(
         \begin{array}{c}
           \delta\psi_i(x) \\
           \delta\psi_i^*(x) \\
         \end{array}\right);\\
         \nonumber
\hat{\mathcal{G}}^{-1}_0=\left(
  \begin{array}{cc}
    \textrm{i}\partial_t-\varepsilon_\textbf{p}-gn_c & -gn_c \\
    -gn_c & -\textrm{i}\partial_t-\varepsilon_\textbf{p}-gn_c \\
  \end{array}
\right),\\
\nonumber
\hat{\mathfrak{G}}^{-1}_0=\left(
  \begin{array}{cc}
    \textrm{i}\partial_t-\varepsilon_\textbf{p}-\Delta & 0 \\
    0 & -\textrm{i}\partial_t-\varepsilon_\textbf{p}-\Delta \\
  \end{array}
\right).
\end{gather}

%--------------------------------------------------------------------------------------
%--------------------------------------------------------------------------------------
%--------------------------------------------------------------------------------------

\textit{Current of the BEC. }Substituting the formal solution of Eq.~\eqref{EqPsi2} into~\eqref{eq.20w}, we come up with an integro-differential Gross--Pitaevskii equation that describes the BEC dynamics accounting for the exciton transitions to the upper state:
\begin{gather}\label{eq.19}
\hat{\mathcal{G}}^{-1}_0\delta\hat{\psi}_1(x)=\left(\frac{e}{mc}\right)^2\textbf{A}(x)\hat{\textbf{q}}\int dx_1\hat{\mathfrak{G}}_0(x-x_1)
\\
\nonumber
~~~~~~~~~~~~~~~~~~~~~~~~~~~~~~~~~\times\textbf{A}(x_1)\hat{\textbf{q}}^*\Bigl(\hat{\psi}_0+\delta\hat{\psi}_1(x_1)\Bigr).
\end{gather}
The corresponding homogeneous system of equations (in the absence of the RHS of Eq.~\eqref{eq.19}), i.e. $\hat{\mathcal{G}}_0^{-1}\delta\hat{\psi}_1(x)=0$, describes the Bogoliubov excitations with the dispersion $\omega_\textbf{p}=sp\sqrt{1+(\xi p)^2}$, where $\xi=1/(2Ms)$ is the healing length and $s=\sqrt{gn_c/M}$ is the sound velocity.  
It can be easily shown that the term containing $\hat\psi_0$ does not contribute to the drag current. As for the term $\delta\hat\psi_1(x_1)$, its contribution can be represented via the Feynman diagrams in Fig.~\ref{Fig2}. 

\begin{figure*}[!t]
\includegraphics[width=0.99\linewidth]{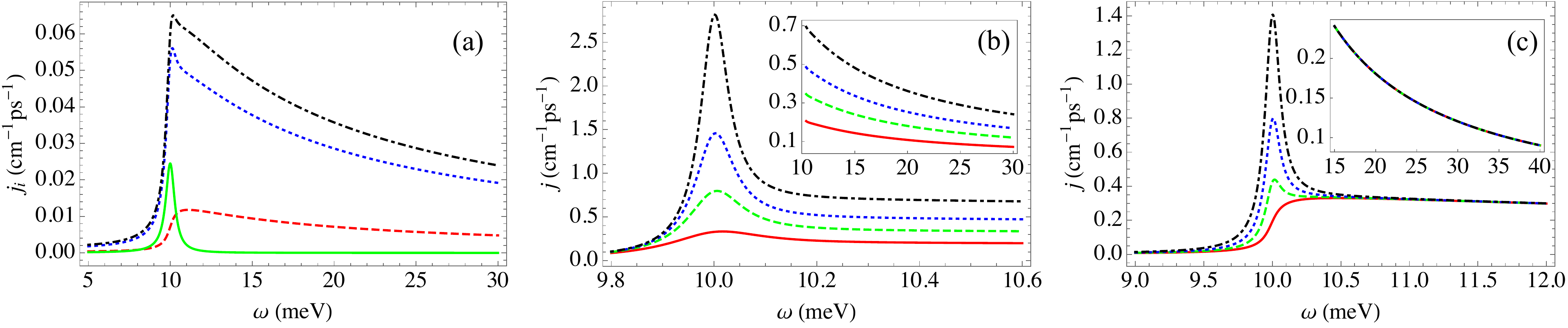}
	\caption{Current as a function of external EMF frequency. (a) Components of current $j_c$ (red), $j_1$ (green), $j_n$ (blue), and full current $j$ (black) for $\tau=1$ ps. (b) Full current for various impurity scattering times: $\tau=3$ ps (red), $5$ ps (green), $7$ ps (blue), and $10$ ps (black). (c) Full current for various BEC concentrations: $n_c=10^{12}$ (red), $2\cdot 10^{13}$ (green), $5\cdot 10^{13}$ (blue), and $10^{14}$ (black) cm$^{-2}$. Insets demonstrate current decay at high $\omega$'s.}
\label{Fig4}
\end{figure*}

We analyze these diagrams in the framework of a linear approximation of the bogolon dispersion, $\omega_\textbf{p}\approx sp$, which is feasible at $\xi p\ll 1$ (see Supplemental Material~\cite{SM}). Accounting for the fact that indirect excitons in a BEC state are more robust against impurity scattering~\cite{KovalevSavenko}, we find:
\begin{eqnarray}\label{EqCurrentBEC}
\textbf{j}_c&=&
\left(\frac{e}{mc}\right)^2
\left|\textbf{q}_{12}\cdot\textbf{A}_0\right|^2
\frac{\tau \textbf{k}}{8\pi\hbar^2}\\
\nonumber
&&\times\Bigl(\arctan\Bigl[2\tau(\omega+\Delta)\Bigr]+\arctan\Bigl[2\tau(\omega-\Delta)\Bigr]\Bigr),
\end{eqnarray}
where $\tau$ is the exciton--impurity scattering time for the non-condensed excitons at the excited state, $|1,\pm 1\rangle$. 
The coefficient containing the vector potential, $\textbf{A}_0$, is connected with the electric field amplitude $\textbf{E}_0$ as
\begin{gather}\label{eq.27w}
\left|\frac{e\textbf{q}_{12}\cdot\textbf{A}_0}{m c}\right|^2=\frac{\Delta^2}{\omega^2}|\textbf{d}_{12}\cdot\textbf{E}_0|^2,
\end{gather}
where $\hat{\textbf{d}}=-e\hat{\textbf{r}}$ is the operator of the in-plane dipole moment of the exciton.

%-----------------------------------
%-----------------------------------
%-----------------------------------

\textit{Current of excited particles in the presence of BEC. }Theoretical derivation of the formula for the current of excited particles ($\delta\hat{\psi}_2$ component), $\textbf{j}_n$, is given in the Supplemental Material~\cite{SM} (see also~\cite{RefOurJETP} for the drag current of indirect excitons in normal state of the gas). 
%%%
It shows that the current consists of two principal parts:
\begin{eqnarray}
\label{EqNormalCurrent}
\textbf{j}_n=\textbf{j}_1+\textbf{j}_2. 
\end{eqnarray}
The first component has a resonant structure as a function of external field frequency,  
\begin{eqnarray}\label{eq.40}
\textbf{j}_1&=&
\left(\frac{e}{mc}\right)^2
\left|\textbf{q}_{12}\cdot\textbf{A}_0\right|^2
\frac{2n_c\textbf{k}\tau^2}{M\hbar}
\\
\nonumber
&&~~~~~~~~~~\times\Bigl[\frac{1}{1+4\tau^2(\omega-\Delta)^2}-\frac{1}{1+4\tau^2(\omega+\Delta)^2}\Bigr], 
\end{eqnarray}
while the second component can be estimated using the Feynman diagrams (Fig.~\ref{Fig2}); direct calculation shows that it satisfies the relation $\textbf{j}_2=4\textbf{j}_c$, where $\textbf{j}_c$ is given in Eq.~\eqref{EqCurrentBEC}. Summing up all the terms, we find the total current of excitons in the system:
\begin{eqnarray}
\label{EqTotalCurrent}
\textbf{j}&=\textbf{j}_c+\textbf{j}_n=5\textbf{j}_c+&\textbf{j}_1.
\end{eqnarray}
%

%----------------------------------
%----------------------------------
%----------------------------------

\textit{Results and discussion. }
The results of calculations are presented in Figs.~\ref{Fig3}--\ref{Fig4}. To build these plots, we used the following parameters: $\Delta=10$ meV (see Supplemental Material~\cite{SM}) and $M=0.5$$m_0$, where $m_0$ is free electron mass. For the matrix element of exciton transition in~\eqref{eq.27w}, we took $|\textbf{d}_{12}\cdot\textbf{E}_0|=0.01\Delta$. Evidently, this value can be controlled by the amplitude of the external EMF as long as $|\textbf{d}_{12}\cdot\textbf{E}_0|\ll\Delta$, since the perturbation theory analysis used here is only legitimate if the external light is not too strong.

Figure~\ref{Fig3} demonstrates the influence of each component constituting total current $\textbf{j}$. Both $j_c$ and $j_n$ increase with exciton--impurity scattering time $\tau$, as expected. One can see that $j_c$ has a clear thresholdlike behavior (processes II in Fig.~\ref{Fig1}b), whereas $j_1$, which is one of the components of $j_n$, has a resonant character (processes I in Fig.~\ref{Fig1}b). 
Let us now address the microscopic nature of each of the components. It is well known that the second-order response functions in a stationary regime can be expressed via the imaginary part of the first-order response functions~\cite{CoulombDrag}, or in other words, by the light absorption coefficient. 
The first term, $\textbf{j}_1$, demonstrating resonant behavior, corresponds to the standard process of light quantum absorption by a BEC particle acompanied by the transition of this particle to the upper excited state, obeying the energy conservation law $\omega=\varepsilon_{\textbf{p}+\textbf{k}}+\Delta$. The second contribution, $\textbf{j}_2$, corresponds to the Belyaev process, which consists of two excitations of the BEC due to the absorption of the EMF quantum, $\omega=\omega_\textbf{p}+\varepsilon_{\textbf{p}+\textbf{k}}+\Delta$, demonstrating thresholdlike dependence on frequency~\cite{Chaplikkovalev} (see Fig.~\ref{Fig1}b, red arrows).

Figure~\ref{Fig4} shows the spectra of the total current in the system. As exhibited in Fig.~\ref{Fig4}a, its shape is a superposition of threshold-like and peaked forms. Total current increases with the increase of impurity scattering time (Fig.~\ref{Fig4}b) since both current components increase, as shown in Fig.~\ref{Fig3}a--b. Total current also grows with the concentration of particles in the BEC, as shown in Fig.~\ref{Fig4}c. 
As expected from Eqs.~\eqref{eq.40} and~\eqref{EqTotalCurrent}, at small $n_c$, the biggest influence on current comes from the $j_2$ term. It should be noted that with the increase of frequency, current vanished when $\omega\gg\Delta$ (see insets in Fig.~\ref{Fig4}b--c). 

All these properties of exciton current allow the proposal of a frequency--selective photodetector employing the effects described above. Indeed, the $j_1$ component of total current---provided that there are a substantial number of particles in the BEC---allows for the suppression of low and high frequencies of the output signal (see Fig.~\ref{Fig4}). Attenuation of such a filter depends first on the concentration of particles in the BEC (see Fig.~\ref{Fig4}c), and second on the purity of the sample, which determines the scattering of particles on impurities through the parameter $\tau$ (see Fig.~\ref{Fig4}b). Since condensates usually have sufficiently low responsiveness to impurities~\cite{KovalevSavenko}, we expect high signal/noise ratio of our detector.
 Here the sensitivity (responsivity) to incoming light is expected to be condensate density-dependent. However, an increase of $n_c$ should not bring any decoherence in the system since the particles are mostly residing one quantum state.
 Important to mention again, comparing with detectors employed in standard electronics, our device deals with the particle density flux (and not with electric current).

%----------------------------------
%----------------------------------
%----------------------------------

\textit{Conclusions. }We have developed a theory of the resonant photon drag effect in a system with indirect excitons in a double quantum well structure under the action of external circularly polarized light. It has been shown that the photon drag flux of excitons experiences resonant behavior when the frequency of light is close to the gap between the ground and excited energy levels of internal exciton motion. The resulting drag current consists of both Bose-condensed and excited particles; as a result, the shape of the total current represents a hybrid of thresholdlike and resonant contributions. These features allow us to propose a photon drag-based frequency--selective photodetector.

The authors would like to thank J. Rasmussen (RECON) for a critical reading of our manuscript. We acknowledge the support by the Russian Science Foundation (Project No. 17-12-01039) and the Institute for Basic Science in Korea (Project No.~IBS-R024-D1).
%------------------------------------------
%------------------------------------------
%------------------------------------------

\pagebreak

%-----------------------------------------
%-----------------------------------------
%-----------------------------------------
%-----------------------------------------
%-----------------------------------------
%-----------------------------------------
%-----------------------------------------

%
\begin{widetext}

\pagebreak

\appendix

\section{SUPPLEMENTAL MATERIAL: Proposal for frequency-selective photodetector based on the resonant photon drag effect in a condensate of indirect excitons}

In this Supplemental Material we provide the derivation of the formulas for the components of the total current and estimation of the energy gap.

%----------------------------------
%----------------------------------
%----------------------------------

\subsection{A. Current in the Bose-Einstein condensate}
\label{ApA}

To find the current, we start with Eq.~(10) in the main text, 
\begin{gather} %\label{eq.21app}
\nonumber
\hat{\mathcal{G}}^{-1}_0\delta\hat{\psi}_1(x)=\left(\frac{e}{mc}\right)^2\textbf{A}(x)\hat{\textbf{q}}\int dx_1\hat{\mathfrak{G}}_0(x-x_1)\textbf{A}(x_1)\hat{\textbf{q}}^*\Bigl(\hat{\psi}_0+\delta\hat{\psi}_1(x_1)\Bigr).
\end{gather}
It is easy to show that the term $\psi_0$ in its r.h.s. gives no contribution to current. Indeed, this term is already of the second order with respect to vector-potential. Thus the current is proportional to $\langle\psi_0\nabla\delta\psi^*\rangle-\langle\delta\psi\nabla\psi_0^*\rangle$, where angular brackets stand for time averaging, which gives zero for the term $\langle\psi_0\nabla\delta\psi^*\rangle$ by definition of fluctuaitons. The term $\langle\delta\psi\nabla\psi_0^*\rangle$ is zero since $\psi_0=\psi_0^*=\sqrt{n_c}$.
Thus, we will further consider the terms containing $\delta\hat{\psi}_1(x')$ only. In the first order in $(\textbf{q}\textbf{A})^2$, we yield:
\begin{gather}\label{eq.22}
\hat{\mathcal{G}}(x,x')=\hat{\mathcal{G}}_0(x-x')+\left(\frac{e}{m c}\right)^2\int dx_1\int dx_2
\hat{\mathcal{G}}_0(x-x_1)\textbf{A}(x_1)\hat{\textbf{q}}
\hat{\mathfrak{G}}_0(x_1-x_2)\textbf{A}(x_2)\hat{\textbf{q}}^*\hat{\mathcal{G}}_0(x_2-x');\\\nonumber
\textbf{j}_c=\frac{1}{2M}\lim_{x'\rightarrow x}\left(\nabla_{\textbf{R}}-\nabla_{\textbf{R}'}\right)\,\langle\hat{\mathcal{G}}^<_{(11)}(x,x')\rangle,
\end{gather}
where $\hat{\mathcal{G}}^<_{(11)}(x,x')$ is the  matrix element of $\hat{\mathcal{G}}^<(x,x')$.
% and $\langle...\rangle$ is time averaging.

The BEC current can be written in the form:
\begin{gather}\label{eq.23}
\textbf{j}_c=\frac{\textrm{i}}{M}\left|\frac{e\textbf{q}_{12}\cdot\textbf{A}_0}{mc}\right|^2\Bigl(\textbf{D}(\textbf{k},\omega)+\textbf{D}(-\textbf{k},-\omega)\Bigr)_{(11)},~~~\textrm{where}\\\nonumber
\hat{\textbf{D}}(\textbf{k},\omega)=\sum_{\textbf{p},\varepsilon}\textbf{p}
\Bigl[\hat{\mathcal{G}}^R_0(\varepsilon+\omega,\textbf{p})\hat{\mathfrak{G}}^R_0(\varepsilon,\textbf{p}-\textbf{k})\hat{\mathcal{G}}_0^<(\varepsilon+\omega,\textbf{p})
+\hat{\mathcal{G}}^R_0(\varepsilon+\omega,\textbf{p})\hat{\mathfrak{G}}^<_0(\varepsilon,\textbf{p}-\textbf{k})\hat{\mathcal{G}}^A_0(\varepsilon+\omega,\textbf{p})+\\\nonumber
+\hat{\mathcal{G}}^<_0(\varepsilon+\omega,\textbf{p})\hat{\mathfrak{G}}^A_0(\varepsilon,\textbf{p}-\textbf{k})\hat{\mathcal{G}}^A_0(\varepsilon+\omega,\textbf{p})\Bigr].
\end{gather}
Here the Green's functions read~\cite{RefKeldysh}:
\begin{gather}\label{eq.24}
\hat{\mathcal{G}}^R_0(\varepsilon,\textbf{p})=\frac{1}{(\varepsilon+\textrm{i}\gamma_{\textbf{p}})^2-\omega_\textbf{p}^2}\left(
                                        \begin{array}{cc}
                                          \varepsilon+\varepsilon_{\textbf{p}}+gn_c & -gn_c \\
                                          -gn_c & -\varepsilon+\varepsilon_{\textbf{p}}+gn_c \\
                                        \end{array}
\right),\\
\nonumber
\hat{\mathcal{G}}^<_0(\varepsilon,\textbf{p})=n_{\textbf{p}}\left[\hat{\mathcal{G}}^R_0(\varepsilon,\textbf{p})-\hat{\mathcal{G}}^A_0(\varepsilon,\textbf{p})\right]
=-\frac{2\pi \textrm{i}}{2\omega_{\textbf{p}}}\left(
                                        \begin{array}{cc}
                                          \varepsilon+\varepsilon_{\textbf{p}}+gn_c & -gn_c \\
                                          -gn_c & -\varepsilon+\varepsilon_{\textbf{p}}+gn_c \\
                                        \end{array}
                                      \right)
\Bigl[n_{\textbf{p}}\delta(\varepsilon-\omega_{\textbf{p}})+[1+n_{\textbf{p}}]\delta(\varepsilon+\omega_{\textbf{p}})\Bigr],\\
\nonumber
\hat{\mathfrak{G}}^R_0(\varepsilon,\textbf{p})=\left(
                                                 \begin{array}{cc}
                                                   \frac{1}{\varepsilon-\varepsilon_\textbf{p}-\Delta+\textrm{i}/2\tau} & 0 \\
                                                   0 & \frac{1}{-\varepsilon-\varepsilon_\textbf{p}-\Delta+\textrm{i}/2\tau} \\
                                                 \end{array}
                                               \right),\\                                               \nonumber
\hat{\mathfrak{G}}^<_0(\varepsilon,\textbf{p})=f_{\textbf{p}}\Bigl[\hat{\mathfrak{G}}_0^R(\varepsilon,\textbf{p})-\hat{\mathfrak{G}}_0^A(\varepsilon,\textbf{p})\Bigr].
\end{gather}
In equations above, $\omega_\textbf{p}=sp\sqrt{1+(\xi p)^2}$ is the dispersion of the Bogoliubov quasi-particles, $\gamma_\textbf{P}=(p\xi)^3/\tau$ is their damping due to scattering on impurities~\cite{KovalevSavenko},  and the distribution functions read

$$f_\textbf{p}=\frac{1}{e^{(\varepsilon_{\textbf{p}}+\Delta-\mu)/T}-1},\,\,\,n_\textbf{P}=\frac{1}{e^{\omega_{\textbf{p}}/T}-1}.$$
At zero temperature, we have $f_\textbf{p}=0,\,n_\textbf{p}=0$. Further we will be assuming linear dispersion of Bogoliubov excitations which holds at $\xi p\ll 1$.

Integrating via $\varepsilon$ in (\ref{eq.23}), we find:
\begin{eqnarray}
\label{eq.25}
\Bigl(\textbf{D}(\textbf{k},\omega)+\textbf{D}(-\textbf{k},-\omega)\Bigr)_{(11)}
&=&
-\textrm{i}\frac{(gn_c)^2}{2}\sum_{\textbf{p}}
\frac{\textbf{p}}{\omega_{\textbf{p}}}\textmd{Re}\,\frac{1}{(\omega_{\textbf{p}}-\textrm{i}\gamma_{\textbf{p}})^2-\omega_{\textbf{p}}^2}
\\
\nonumber
&&\times
\left[-\frac{1}{\omega+\omega_\textbf{p}+\varepsilon_{\textbf{p}-\textbf{k}}+\Delta-\textrm{i}/2\tau}
+\frac{1}{\omega-\omega_\textbf{p}-\varepsilon_{\textbf{p}+\textbf{k}}-\Delta+\textrm{i}/2\tau}
\right.
\\
\nonumber
&&~~~~~~~~\left.+\frac{1}{\omega+\omega_\textbf{p}-\varepsilon_{\textbf{p}-\textbf{k}}-\Delta+\textrm{i}/2\tau}
-\frac{1}{\omega-\omega_\textbf{p}+\varepsilon_{\textbf{p}+\textbf{k}}+\Delta-\textrm{i}/2\tau}
\right].
\end{eqnarray}
Extracting the real and imaginary parts of the expression in square brackets in~\eqref{eq.25} and assuming $
\omega_\textbf{p}\gg\varepsilon_{\textbf{p}\pm \textbf{k}}$, for small $\textbf{k}$ we find:
\begin{eqnarray}
\label{eq.26}
\Bigl(\textbf{D}(\textbf{k},\omega)+\textbf{D}(-\textbf{k},-\omega)\Bigr)_{(11)}
&=&
-\textrm{i}\frac{(gn_c)^2}{2M}\sum_{\textbf{p}}
\frac{\textbf{p}}{\omega_{\textbf{p}}}\frac{\textbf{k}\cdot\textbf{p}}{(4\omega^2_\textbf{p}+\gamma^2_\textbf{p})}
\\
\nonumber
&&\times
\left[-\frac{1}{(\omega+\Delta+\omega_\textbf{p})^2+1/4\tau^2}+\frac{1}{(\omega+\Delta-\omega_\textbf{p})^2+1/4\tau^2}\right.
\\
\nonumber
&&~~~~~~~~\left.-\frac{1}{(\omega-\Delta+\omega_\textbf{p})^2+1/4\tau^2}+\frac{1}{(\omega-\Delta-\omega_\textbf{p})^2+1/4\tau^2}\right].
\end{eqnarray}
Further taking into account that $\omega_\textbf{p}\gg\gamma_\textbf{p}$ and integrating over $\textbf{p}$, we find:
\begin{gather}\label{eq.27}
\Bigl(\textbf{D}(\textbf{k},\omega)+\textbf{D}(-\textbf{k},-\omega)\Bigr)_{(11)}=-\textrm{i}\frac{(gn_c)^2\tau \textbf{k}}{8\pi Ms^4}\Bigl[\arctan\Bigl(2\tau(\omega+\Delta)\Bigr)+\arctan\Bigl(2\tau(\omega-\Delta)\Bigr)\Bigr].
\end{gather}
Remembering $gn_c=Ms^2$, we finally find the current of BEC particles:
\begin{gather}\label{eq.28}
\textbf{j}_c=\left|\frac{e\textbf{q}_{12}\cdot\textbf{A}_0}{mc}\right|^2\frac{\tau \textbf{k}}{8\pi\hbar^2}\Bigl(\arctan\Bigl[2\tau(\omega+\Delta)\Bigr]+\arctan\Bigl[2\tau(\omega-\Delta)\Bigr]\Bigr).
\end{gather}
%

%----------------------------
%----------------------------
%----------------------------
%----------------------------
%----------------------------

\subsection{B. Current of excited particles in the presence of a BEC}
\label{ApB}

Expressing $\delta\hat{\psi}_1(x)$ via $\delta\hat{\psi}_2(x)$,
\begin{gather}\label{eq.29}
\delta\hat{\psi}_1(x)=-\frac{e}{mc}\int dx_1\hat{\mathcal{G}}_0(x-x_1)\textbf{A}(x_1)\hat{\textbf{q}}\delta\hat{\psi}_2(x_1),
\end{gather}
we can find the closed system of equations for $\delta\psi_2(x)$:
\begin{gather}\label{eq.30}
\mathfrak{G}^{-1}_0\delta\hat{\psi}_2(x)=-\frac{e}{mc}\textbf{A}^*(x)\hat{\textbf{q}}^*\Bigl(\hat{\psi}_0-\frac{e}{mc}\int dx_1\hat{\mathcal{G}}_0(x-x_1)\textbf{A}(x_1)\hat{\textbf{q}}\delta\hat{\psi}_2(x_1)\Bigr).
\end{gather}
Here in the r.h.s. we have two contributions, one containing $\textbf{A}$ and the other containing $|\textbf{A}|^2$ terms. Let us analyze them separately.

%----------------------------

\subsubsection{B1. Contribution of the first order in $\textbf{A}$ term}

Keeping the $\textbf{A}$ term in (\ref{eq.30}) only and making the Fourier transform, we come up with the wave function:
\begin{gather}\label{eq.31}
\delta\hat{\psi}_2(x)=-\frac{e}{m c}\Bigl[\hat{\mathfrak{G}}_0(\textbf{k},\omega)\textbf{A}_0\hat{\textbf{q}}\hat{\psi}_0e^{\textrm{i}\textbf{kr}-\textrm{i}\omega t}+\hat{\mathfrak{G}}_0(-\textbf{k},-\omega)\textbf{A}^*_0\hat{\textbf{q}}^*\hat{\psi}_0e^{-\textrm{i}\textbf{kr}+\textrm{i}\omega t}\Bigr],
\end{gather}
and then using Eq.~(6) from the main text and similar derivation as in Appendix~A above, we can find an expression for the current:
\begin{gather}\label{eq.31}
\textbf{j}_1=\frac{2n_c\textbf{k}\tau^2}{M\hbar}\left|\frac{e\textbf{q}_{12}\cdot\textbf{A}_0}{mc}\right|^2
\Bigl[\frac{1}{1+4\tau^2(\omega-\Delta)^2}-\frac{1}{1+4\tau^2(\omega+\Delta)^2}\Bigr].
\end{gather}
%

%------------------------------------

\subsubsection{B2. Contribution of the second order in $\textbf{A}$ term}
Algebraic analysis of the $|\textbf{A}|^2$ term in (\ref{eq.30}) gives the following expression for the current:
\begin{gather}\label{eq.34}
\textbf{j}_2=\frac{\textrm{i}}{M}\left|\frac{e\textbf{q}_{12}\cdot\textbf{A}_0}{m c}\right|^2\Bigl[ \hat{\mathcal{D}}(\textbf{k},\omega)+\hat{\mathcal{D}}(-\textbf{k},-\omega)\Bigr]_{(11)},~\textrm{where}\\
\nonumber
\hat{\mathcal{D}}(\textbf{k},\omega)=\sum_{\textbf{p},\varepsilon}\textbf{p}
\Bigl[\hat{\mathfrak{G}}^R_0(\varepsilon+\omega,\textbf{p})\hat{\mathcal{G}}^R_0(\varepsilon,\textbf{p}-\textbf{k})\hat{\mathfrak{G}}_0^<(\varepsilon+\omega,\textbf{p})
+\hat{\mathfrak{G}}^R_0(\varepsilon+\omega,\textbf{p})\hat{\mathcal{G}}^<_0(\varepsilon,\textbf{p}-\textbf{k})\hat{\mathfrak{G}}^A_0(\varepsilon+\omega,\textbf{p})+\\\nonumber
+\hat{\mathfrak{G}}^<_0(\varepsilon+\omega,\textbf{p})\hat{\mathcal{G}}^A_0(\varepsilon,\textbf{p}-\textbf{k})\hat{\mathfrak{G}}^A_0(\varepsilon+\omega,\textbf{p})\Bigr].
\end{gather}
Using (\ref{eq.24}) at zero temperature, when $f_\textbf{p}=n_\textbf{p}=0$, we find:
\begin{eqnarray}
\label{eq.35}
&&\Bigl(\mathcal{D}(\textbf{k},\omega)+\mathcal{D}(-\textbf{k},-\omega)\Bigr)_{(11)}\\
\nonumber
&&~~~~~~~~=-\textrm{i}\frac{gn_c}{2}
\sum_{\textbf{p}}\left[\frac{\textbf{p}}{\omega_{\textbf{p}-\textbf{k}}}
\frac{1}{(\omega-\omega_{\textbf{p}-\textbf{k}}-\varepsilon_{\textbf{p}}-\Delta)^2+1/4\tau^2}+
\frac{\textbf{p}}{\omega_{\textbf{p}+\textbf{k}}}
\frac{1}{(-\omega-\omega_{\textbf{p}+\textbf{k}}-\varepsilon_{\textbf{p}}-\Delta)^2+1/4\tau^2}\right]\\
\nonumber
&&~~~~~~~~=-\textrm{i}\frac{gn_c}{2}
\sum_{\textbf{p}}\left[\frac{\textbf{p}+\textbf{k}}{\omega_{\textbf{p}}}
\frac{1}{(\omega-\omega_{\textbf{p}}-\varepsilon_{\textbf{p}+\textbf{k}}-\Delta)^2+1/4\tau^2}+
\frac{\textbf{p}-\textbf{k}}{\omega_{\textbf{p}}}
\frac{1}{(\omega+\omega_{\textbf{p}}+\varepsilon_{\textbf{p}-\textbf{k}}+\Delta)^2+1/4\tau^2}\right].
\end{eqnarray}
If we now take into account that $\omega_\textbf{p}\gg\varepsilon_{\textbf{p}\pm \textbf{k}}$, we come up with
\begin{gather}\label{eq.36}
\Bigl(\mathcal{D}(\textbf{k},\omega)+\mathcal{D}(-\textbf{k},-\omega)\Bigr)_{(11)}=-\textrm{i}\frac{gn_c\textbf{k}}{2}
\sum_{\textbf{p}}\frac{1}{\omega_{\textbf{p}}}\left[
\frac{1}{(\omega-\omega_{\textbf{p}}-\Delta)^2+1/4\tau^2}+
-\frac{1}{(\omega+\omega_{\textbf{p}}+\Delta)^2+1/4\tau^2}\right].
\end{gather}
Finally, integrating over $\textbf{p}$ in~\eqref{eq.36} and taking into account that $gn_c=Ms^2$, we find the resulting equation for the current density:
\begin{gather}\label{eq.37}
\textbf{j}_2=\left|\frac{e\textbf{q}_{12}\cdot\textbf{A}_0}{mc}\right|^2\frac{\tau \textbf{k}}{2\pi\hbar^2}\Bigl(\arctan\Bigl[2\tau(\omega+\Delta)\Bigr]+\arctan\Bigl[2\tau(\omega-\Delta)\Bigr]\Bigr).
\end{gather}
It is important to note that, comparing Eq.~\eqref{eq.37} with~\eqref{eq.28}, we find $\textbf{j}_2=4\textbf{j}_c$. Combining together Eq.~\eqref{eq.37} with~\eqref{eq.31}, we can find the current of excitons in the normal state defined in Eq.~(13) in the main text.

%----------------------------
%----------------------------
%----------------------------

\subsection{C. Energy gap $\Delta$}

Here we estimate the value of $\Delta$ for different materials. As an example, let us consider double quantum wells based on the alloy GaAs and heterojunctions based on MoS$_2$. The interaction energy in the former case reads
\begin{gather}\label{eq.C1}
U^{(1)}_c(\textbf{r})=-\frac{e^2}{\varepsilon_d\sqrt{a^2+r^2}},
\end{gather}
and in the latter case it is
\begin{gather}\label{eq.C2}
U^{(2)}_c(\textbf{r})=-\frac{\pi e^2}{2\varepsilon_d\rho_0}\Bigl[H_0\left(\frac{\sqrt{a^2+r^2}}{\rho_0}\right)-Y_0\left(\frac{\sqrt{a^2+r^2}}{\rho_0}\right)\Bigr],
\end{gather}
where $\varepsilon_d$ is a dielectric function of the media, $\rho_0=2\pi\alpha/\varepsilon_d$, where $\alpha$ is 2D polarizability; $H_0$, $Y_0$ are zero-order Struve and Bessel functions, respectively. In the limiting case $r\ll a$, both the functions can be written as:
\begin{gather}\label{eq.C3}
U_c(\textbf{r})\approx C+\frac{m\omega_0^2r^2}{2},
\end{gather}
where $C$ is a constant. Equation~\eqref{eq.C3} was found as an expansion of~\eqref{eq.C1} and~\eqref{eq.C2} over a small parameter $r/a$, which gives
\begin{gather}\label{eq.C4}
{\omega_0^{(1)}}^2=\frac{e^2}{\varepsilon_dm a^3},\\\nonumber
{\omega_0^{(2)}}^2=-\frac{\pi e^2}{2\varepsilon_dm a\rho_0^2}
\Bigl[H_{-1}\left(\frac{a}{\rho_0}\right)-Y_{-1}\left(\frac{a}{\rho_0}\right)\Bigr]
\end{gather}
for GaAs and MoS$_2$, correspondingly. Using standard parameters and assuming $\Delta\approx\omega_0$, we can estimate $\Delta$ as $\sim 10$ meV for GaAs and $\sim 40$ meV for MoS$_2$.

\end{widetext}

\end{document}